\documentclass[letter]{aa}

\usepackage{graphicx,epsfig}

\newcommand{\fgas}{f_{\rm gas}}

\newcommand{\Ho}{H_{\rm o}}
\newcommand{\OmM}{\Omega_{\rm M}}
\newcommand{\OmL}{\Omega_\Lambda}

\newcommand{\thetac}{\theta_{\rm c}}
\newcommand{\tauc}{\tau_{\rm c}}
\newcommand{\thetaco}{\theta_{\rm co}}
\newcommand{\Yo}{Y_{\rm o}}

\newcommand{\sigN}{\sigma_{\rm N}}
\newcommand{\fwhm}{\theta_{\rm fwhm}}

\newcommand{\No}{N_{\rm o}}
\newcommand{\Mdet}{M_{\rm det}}

\begin{document}
   \title{The Selection Function of SZ Cluster Surveys}

   \subtitle{}

   \author{J.--B. Melin
          \and J.G. Bartlett 
	  \and J. Delabrouille
          }

   \offprints{J.--B. Melin}

   \institute{APC -- Universit{\'e} Paris 7, Paris, France\\
             \email{melin@cdf.in2p3.fr, bartlett@cdf.in2p3.fr, delabrouille@cdf.in2p3.fr}
             }

   \date{}

   \abstract{We study the nature of cluster selection in
   Sunyaev-Zel'dovich (SZ) surveys, focusing on single frequency
   observations and using Monte Carlo simulations incorporating
   instrumental effects, primary cosmic microwave background (CMB)
   anisotropies and extragalactic point sources.  Clusters are
   extracted from simulated maps with an optimal, multi--scale matched
   filter.  We introduce a general definition for the survey selection
   function that provides a useful link between an observational
   catalog and theoretical predictions.  The selection function
   defined over the observed quantities of flux and angular size is
   independent of cluster physics and cosmology, and thus provides a
   useful characterization of a survey.  Selection expressed in terms
   of cluster mass and redshift, on the other hand, depends on both
   cosmology and cluster physics.  We demonstrate that SZ catalogs are
   not simply flux limited, and illustrate how incorrect modeling of
   the selection function leads to biased estimates of cosmological
   parameters.  The fact that SZ catalogs are not flux limited
   complicates survey ``calibration'' by requiring more detailed
   information on the relation between cluster observables and cluster
   mass.  }

   \maketitle


\section{Introduction}

Galaxy cluster surveys are important tools for measuring key
cosmological quantities and for understanding the process of structure
formation in the universe (\cite{bah99}; \cite{ros02}). Surveying for
clusters using the Sunyaev--Zel'dovich (SZ) effect (\cite{sun70};
\cite{sun72}; for recent reviews, see \cite{bir99}, and \cite{car02}) offers 
a number of advantages over more traditional methods based on X--ray
or optical imaging.  These advantages include good detection
efficiency at high--redshift; a selection based on the thermal energy
of the intracluster medium, a robust quantity relative to any thermal
structure in the gas; and an almost constant mass detection limit with
redshift (\cite{hol00}; \cite{bar00}; \cite{bar01}).  A new generation
of optimized, dedicated instruments, both large bolometer arrays
(\cite{mas03}; \cite{run03}; \cite{kos04}) and interferometers
(\cite{lo00}; \cite{jones02}), will soon perform such SZ cluster
surveys, and we may look forward to the large and essentially
full--sky SZ catalog expected from the Planck mission\footnote{A list of web
pages describing a number of experiments is given in the reference
section.}.
In anticipation, many authors have studied
the nature and use of SZ cluster catalogs and made predictions for the
number of objects expected from various proposed surveys
(\cite{hol00}; \cite{kne01}).  A good example of the potential of an
SZ survey is the use of its redshift distribution to examine structure
formation at high redshift and to thereby constrain cosmological
parameters, such as the density parameter $\OmM$ (\cite{barb96}), 
and the dark energy equation--of--state $\omega$ (\cite{haiman01}).

An astronomical survey is fundamentally characterized by its selection
function, which identifies the subclass of objects detected among all
those actually present in the survey area.  It is a function of
cluster properties and survey conditions.  Depending on the nature of
the observations, relevant cluster properties may include: mass,
redshift, luminosity, morphology, etc..., while key descriptors of the
survey would be sensitivity, angular resolution, spectral coverage,
etc...  The selection function will also depend on the the detection
algorithm used to find clusters in the survey data.
Understanding of the selection function is a prerequisite to any 
statistical application of the survey catalog; otherwise, one has no
idea how representative the catalog is of the parent 
population actually out in the universe.  

Selection function issues for SZ surveys have been touched on recently
by several authors (\cite{bar01}; \cite{sch03}; \cite{whi03}), while
most previous studies of the potential use of SZ surveys have not
examined this point in detail.  For example, predictions of the
redshift distribution of SZ--detected clusters
usually assume that they are point sources, simply selected on their
total flux\footnote{The term {\em flux} does not really apply in the
case of SZ observations, as the effect is measured relative to the
unperturbed background and may be negative.  We shall nevertheless use
it throughout for simplicity.}.  We shall see below that this is not
necessarily the case, and an analysis of cosmological parameters based
on such an assumption would significantly bias the results.

Understanding a survey selection function is difficult. By its very
nature and purpose, the selection function is supposed to tell us
about objects that we {\em don't see} in the survey!  Realistic
simulations of a survey are central to determining its selection
function (e.g., \cite{adami01}).  One knows which objects are put into
the simulation and can then compare them to the subset of objects
detected by the mock observations.
In practice, of course, understanding of a selection function comes
only from a combination of such simulations and diverse observations
taken under different conditions and/or in different wavebands; full
understanding thus comes slowly.

There are really two distinct issues connected to the selection
function: object detection, or {\em survey completeness}, and object
measurement, which we shall refer to as {\em photometry}; as a
separate issue, one must also determine the contamination function.
One would like to characterize each detected cluster by determining,
for example, its total flux, angular size, etc...  As practitioners are
well aware, photometry of extended objects faces many difficulties
that introduce additional uncertainty and, in particular, potential
bias into the survey catalog.  The selection function must correct for
bias induced by both the detection and photometric procedures.  The
two are, however, distinct steps in catalog construction, and the
selection function (see below) should reflect this fact.

The object of the present work is to begin a study of SZ selection
functions for the host of SZ surveys that are being planned, and to
propose a formalism for their characterization.  To this end, we have
developed a rapid Monte Carlo simulation tool (\cite{del02}) that
produces mock images of the SZ sky, including various clustering and
velocity effects, primary cosmic microwave background (CMB)
anisotropies, radio point sources and instrumental effects.  The main
goals of such studies, in this period before actual surveying has
begun, are to improve understanding of the expected scientific return
of a given survey and to help optimize observing strategies.

Our specific aim in the present work is to study selection effects in
SZ surveys by focusing on single frequency observations, such as will
be performed by up--coming interferometers.  Most bolometer cameras
propose surveys at several frequencies, although not necessarily
simultaneously; the present considerations are therefore applicable to
the first data sets from these instruments.  This work builds on that
of \cite{bar00} by adding the effects of primary CMB
anisotropies, point sources and photometric errors, and by the use of
an optimized cluster detection algorithm (\cite{melin04}).

General considerations concerning the selection function are given in
the next section and used to motivate our definition given in
Eq.(\ref{eq:selfundef}).  We then briefly describe
(Section~\ref{sec:sims}) our simulations, based on a Monte Carlo
approach incorporating cluster correlations and velocities, as well as
our cluster detection and photometry algorithms built on an optimized
spatial filter (details will be given elsewhere [\cite{melin04}]).  A
discussion of cluster selection with this method follows
(Section~\ref{sec:selfun}), where with a simple analytic argument, we
show how cluster detection depends on {\em both} total flux and
angular size.  Our main conclusion is that SZ surveys will not be
simply flux limited.  Our simulations support the analytical
expectations, and they also highlight the difficulty of performing
accurate photometry on detected clusters.  

We close with a discussion (Section~\ref{sec:dis}) of some
implications for upcoming surveys.  The most important is that the
redshift distribution of observed clusters differs from that of a pure
flux--limited catalog; assuming pure flux selection will therefore
lead to biased estimates of cosmological parameters.  In this same
section, we give an explicit example of biased parameter estimation
caused by the presence of incorrectly modeled excess primary CMB power
on cluster scales, as suggested by the CBI experiment
(\cite{mason01}).  We note that non-trivial cluster selection
complicates survey ``calibration'' (\cite{bartel01};
\cite{hu03}; \cite{maj03}; \cite{lima04}) because a {\em size--mass} relation 
must be obtained in addition to a {\em flux--mass} relation.
Photometric errors will further increase the difficulty by augmenting
scatter in the mass--observable relations.

\section{Selection Function: general considerations}

To motivate our definition, we first consider some general properties
desired of a survey selection function.  Fundamentally, it relates
observed catalog properties (e.g., flux and size) to relevant
intrinsic characteristics of the source population under study.  In
particular, we want it to tell us about the {\em completeness} of the
survey catalog as a function of source properties, which is a measure
of the selection {\em bias}.  In addition, we also wish for it to
reflect the effects of statistical (e.g., photometric) errors.
Notice, on the other hand, that the selection function will not tell
us anything about {\em contamination} of the catalog by false
detections; this is another function of observed quantities that must
be separately evaluated.

Consider the example of a flux--limited catalog of point sources.
Neglecting photometric measurement errors, the probability that a
source at redshift $z$ will find its way into the survey catalog is
simply given by the fraction of sources brighter than the flux limit,
which may be calculated as an integral over the luminosity function at
$z$ (e.g., \cite{peebles93}).  Extended objects complicate the situation,
for their detection will in general depend on morphology.  One must
then define appropriate source descriptors other than just a total flux;
and even the definition of total flux, conceptually simple, becomes
problematic (fixed aperture flux, isophotal flux, integrated flux with
a fitted profile, etc...).  The choice of descriptors is clearly
important and the selection function will depend on it.  They must
encode relevant observational information on the sources and
represent observables with as little measurement error as possible.

The simplest characterization for extended SZ sources would employ a
total observed flux, $\Yo$, and a representative angular size, which
we take to be the core radius $\thetaco$.  By total flux, we mean the
flux density integrated over the entire cluster profile, out to the
virial radius, and we express it in a frequency independent manner as
the integrated Compton--y parameter.  We limit ourselves to these two
descriptors in the ensuing discussion, although clearly many others
describing cluster morphology are of course possible (ellipticity, for
example...).  How the observed quantities are actually measured is
crucial -- measurement errors and the selection function will both
depend on the technique used.

Our detected clusters will then populate the observed parameter space
according to some distribution $d\No/d\Yo d\thetaco$. What we
really seek, however, is the true cluster distribution, $dN/dY
d\thetac$, over the intrinsic cluster parameters $Y$ and $\thetac$.
Measurement errors and catalog incompleteness both contribute to the
difference between these two distributions.  In addition, the catalog
will suffer from contamination by false detections.

These general considerations motivate us to define the {\em selection
function} as the {\em joint distribution of $\Yo$ and $\thetaco$}, as
a function of (i.e., given) $Y$ and $\thetac$.  There are many other
factors that influence the selection function, such as instrument
characteristics, observation conditions and analysis methods, so in
general we write
\begin{equation}\label{eq:selfundef}
\Phi\left[\Yo,\thetaco |Y, \thetac, \sigN, \fwhm, ...\right]
\end{equation}
where $\fwhm$ is the FWHM of an assumed Gaussian beam and
$\sigN^2$ is the map noise variance.  We illustrate our main 
points throughout this discussion with simple uniform Gaussian white noise.  
The dots represent other possible influences on the selection function,
such as the detection and photometry algorithms employed to construct
the catalog. 

Several useful properties follow from this definition.  For example,
the selection function relates the observed counts from a survey to
their theoretical value by
\begin{eqnarray}
\nonumber
\frac{d\No}{d\Yo d\thetaco}\left(\Yo,\thetaco\right) 
	& = &\int_0^\infty dY\; \int_0^\infty d\thetac\; 
	\Phi(\Yo,\thetaco|Y, \thetac)\\
& & 	\times \frac{dN}{dY d\thetac}\left(Y,\thetac\right)
\end{eqnarray}
A similar relation can be established between the 
observed counts and cluster mass and redshift:
\begin{eqnarray}
\nonumber
\frac{d\No}{d\Yo d\thetaco}\left(\Yo,\thetaco\right) 
	& = & \int_0^\infty dz\; \int_0^\infty dM\; 
	\Psi(\Yo,\thetaco|z,M)\\
& & 	\times \frac{dN}{dz dM}\left(z,M\right)
\label{eq:obs2Mz}
\end{eqnarray}
where $dN/dzdM$ is the mass function and $\Psi$ incorporates
the intrinsic and observational scatter in the relation
between $(\Yo,\thetaco)$ and $(z,M)$ (mass--observable relations).  
This is made more explicit by
\begin{eqnarray}
\nonumber
\Psi(\Yo,\thetaco|z,M) &=& \int_0^\infty dY \int_0^\infty d\thetac
	\Phi(\Yo,\thetaco|Y, \thetac)\\
& & 	\times T(Y,\thetac|z,M)
\label{eq:psifunc}
\end{eqnarray}
where the function $T$ represents the intrinsic scatter in
the relation between actual flux $Y$ and core radius $\thetac$,
and cluster mass and redshift. 

In general, we may separate the selection function into two parts, one
related to detection and the other to photometry:
\begin{equation}\label{eq:sfsep}
\Phi(\Yo,\thetaco|Y, \thetac)  = \chi(Y, \thetac)F(\Yo,\thetaco|Y, \thetac)
\end{equation}
The first factor represents survey completeness and is 
simply the ratio of detected to actual clusters as a function
of true cluster parameters. The second factor 
quantifies photometric errors with a distribution function
$F$ normalized to unity:
\begin{displaymath}
\int d\Yo d\thetaco \; F(\Yo,\thetaco|Y, \thetac) = 1
\end{displaymath}
In the absence of measurement errors we would have 
\begin{displaymath}
\Phi(\Yo,\thetaco|Y,\thetac)
= \chi(\Yo,\thetaco)\delta(\Yo-Y)\delta(\thetaco-\thetac)
\end{displaymath}
in which case the observed counts become
\begin{equation}
\label{eq:nodispersion}
\frac{d\No}{d\Yo d\thetaco}\left(\Yo,\thetaco\right) = 
	\chi(\Yo,\thetaco) \frac{dN}{dY d\thetac}\left(Y,\thetac\right)
\end{equation}

The importance of the selection function for cosmological studies lies
in Eq.~(\ref{eq:obs2Mz}) which relates the cosmologically sensitive mass
function to the observed catalog distribution.  Accurate knowledge of
$\Psi$ is required in order to obtain constraints on cosmological
parameters, such as the density parameter or the dark energy
equation--of--state.

\section{Simulations}
\label{sec:sims}
Detailed study of SZ selection issues requires realistic simulations
of proposed surveys.  Although analytic arguments do provide
significant insight, certain effects, such as cluster--cluster
blending and confusion, can only be fully modeled with simulations.
To this end, we have developed a rapid Monte Carlo--based simulation
tool that allows us to generate a large number of realizations of a
given survey.  This is essential in order to obtain good measures of
the selection function that are not limited by insufficient
statistics.  In this section we briefly outline our simulation method
and our cluster detection algorithm, leaving details to \cite{del02}
and \cite{melin04}.  

Unless explicitly stated, the simulations used in this work are for a
flat concordance model (\cite{spe03}) with $\OmM=0.3=1-\OmL$, Hubble
constant of $\Ho = 70$~km/s/Mpc (\cite{hstkey}) and a power spectrum
normalization $\sigma_8=0.98$.  The normalization of the $M-T$
relation is chosen to reproduce the local abundance of X--ray clusters
with this value of $\sigma_8$ (\cite{pierp01}).  Finally, we fix the
gas mass fraction at $\fgas=0.12$ (e.g., \cite{mohr99}).

\subsection{Method}

Our simulations produce sky maps at different frequencies and include
galaxy clusters, primary CMB anisotropies, point sources and
instrumental properties (beam smoothing and noise).  In this work, 
we do not consider diffuse Galactic foregrounds, such as dust and
synchrotron emission, as we are interested in more rudimentary
factors influencing the selection function; we leave foreground
issues to a future work (as general references, see \cite{bougis99}; 
\cite{teg00}; \cite{del03}).

We model the cluster population using the Jenkins et al. (2001) mass 
function and self--similar, isothermal $\beta$--profiles for the SZ
emission.  A realization of the linear density field $\delta\rho/\rho$
within a comoving 3D box, with the observer placed at one end, 
is used to construct the cluster spatial distribution and velocity
field.  We scale the density field by the linear growth factor
over a set of redshift slices (or bins) along the past light--cone
of the observer; a set of mass bins is defined within each redshift 
slice.  We then construct a random cluster catalog by drawing 
the number of clusters in each bin of  mass and redshift according
to a Poisson distribution with mean given by the mass function
integrated over the bin.  Within each redshift slice, we spatially
distribute these clusters with a probability proportional to 
$1+b\frac{\delta\rho}{\rho}$, where $b$ is the linear bias given
by \cite{mo96}.  Comparison of the resulting spatial and velocity
2--point functions of the mock catalog with results from the VIRGO
consortium's N--body simulations shows that this method faithfully
reproduces the correlations down to scales of order of 10$h^{-1}$~Mpc.

Individual clusters are assigned a temperature using a $M-T$ relation
consistent with the chosen value of $\sigma_8$ (\cite{pierp01})  
\begin{equation}
{M \over 10^{15} h^{-1} M_\odot} = \left ( {T \over \beta_p} \right
)^{3 \over 2} \left ( \Delta_c E^2 \right )^{- {1 \over 2}}
\end{equation}
with $\beta_p = 1.3 \pm 0.13 \pm 0.13$~keV.
Here, $\Delta_c$ is the mean density contrast for virialization
(weakly dependent on the cosmology) and $E(z)=H(z)/H_0$.
As mentioned, we distribute the cluster gas with an isothermal
$\beta$--model:
\begin{equation}
 {n_e(r)} = {n_e(0)} { \left [1 + \left ({r \over r_c} \right )^2 \right ]^{-3 \beta \over 2}}
\end{equation}
where we fix $\beta=2/3$ and the core radius is taken to be \mbox{$r_c
= 0.1 \, r_v$}, with the virial radius given by
\begin{equation}
 {r_v} =  { 1.69 \, h^{-2/3} \; \left ( {M \over 10^{15} M_\odot}
  \right )^{1/3} \; \left ( {\Delta_c \over 178} \right )^{-1/3} \; E^{-2/3} \;
  {\rm Mpc}}
\end{equation}
The central electron density is determined by the gas mass fraction
$\fgas$.  For the present work, we ignore any intrinsic scatter in
these scaling relations.

In this way we produce a $3\times 3$ degree map of the SZ sky.
Primary CMB anisotropies are added as a Gaussian random field by
drawing Fourier modes according to a Gaussian distribution with zero
mean and variance given by the power spectrum as calculated with
CMBFAST (\cite{sel96}).  We then populate the maps with radio and
infrared point sources, using the counts summarized in \cite{ben03} and
fitted by \cite{knox03}, and the counts from SCUBA (\cite{borys03}).
Finally, the map is smoothed with a Gaussian beam and
white Gaussian noise is added to model instrumental effects.

\subsection{Detection Algorithm}

We have developed (\cite{melin04}) a rapid detection routine
incorporating a deblending algorithm that is based on matched
filtering (\cite{hae96}), for single frequency surveys, and matched
multi--filtering (\cite{her02}), for multi--frequency surveys.  Recall
that in this work we only examine single frequency surveys.  The
matched filter, on a scale $\theta_c$, is defined to yield the best
linear estimate of the amplitude of the SZ signal from a cluster with
(matched) core radius $\theta_c$.  It depends on both the
beam--smoothed cluster profile $\tauc$ and the noise power spectrum
$P(k)$.  In Fourier space it is given by
\begin{equation}
{\hat F}({\bf k}) = \left [ \int {| {\hat \tauc}({\bf k'}) |^2 \over P(k')} {d^2k' \over (2 \pi)^2} \right ]^{-1} {{\hat \tauc^*}({\bf k}) \over P(k)}
\end{equation}
where $P=(P_{cmb}+P_{sources})|{\hat B}|^2+P_{ins}$, ${\hat
\tauc}$ is the Fourier transform of the beam--smoothed cluster profile 
$\tauc$, ${\hat B}$ is that of the instrumental beam (a Gaussian), and
$P_{cmb}$, $P_{sources}$ and $P_{ins}$ represent the power spectra of
the primary CMB anisotropies, residual point sources and instrumental
noise, respectively.  We denote the standard deviation of the noise 
(including primary CMB and residual points sources)
passed through the filter at scale $\theta_c$ by $\sigma_{\theta_{\rm
c}}$, and give its expression for future reference:
\begin{equation}\label{eq:filternoise}
  \sigma_{\theta_c} =  \left [ \int {|{\hat \tauc}({\bf k})|^2
  \over P(k)} {d^2k \over (2 \pi)^2} \right ]^{-{1 \over 2}}
\end{equation}
This is the fluctuation amplitude of the filtered signal in the
absence of any cluster signal.

We can summarize the detection algorithm in three steps:
\begin{itemize}
\item Filter the observed map with matched filters on different
scales $\thetac$ in order to identify clusters of 
different sizes.  This produces a set of filtered maps.
\item In each filtered map, find the pixels that satisfy
${S \over N} > threshold$ (e.g. 3 or 5).  Define cluster
candidates as local maxima among these pixels.  At this
point, each cluster candidate -- in each map -- has a position, size
(that of the filter that produced the map), and a SZ flux 
given by the signal through the matched filter.
\item Identify cluster candidates across the different filtered maps
using a tree structure (the same cluster can obviously be 
detected in several filtered maps) and eliminate multiple 
detections by keeping only cluster properties corresponding to 
the highest S/N map for each candidate.
\end{itemize} 

\section{Selection Function for Single Frequency SZ Surveys}
\label{sec:selfun}

We consider a single frequency SZ survey with characteristics
representative of upcoming interferometers (e.g., the {\sl Arcminute
MicroKelvin Imager} being constructed in Cambridge\footnote{\tt
http://www.mrao.cam.ac.uk/telescopes/ami/index.html}): a 15~GHz
observation frequency, 2~arcmin FWHM (synthesized) beam and a noise
level of $5\; \mu$K/beam.  Note that, for simplicity and generality, we
model the observations as a fully sampled sky map instead of
actual visibilities.  This approximation should be reasonably
accurate given the good sampling expected in the Fourier plane; it will,
however, miss important details of the selection function that will
require adequate modeling when the time comes.  In the same spirit,
we also model the noise as a white Gaussian random variable with 
zero mean and the given variance. 

During the course of the discussion, we will often compare the
following observational cases: 1) no instrumental noise 
(CMB+beam\footnote{Note that in this case of no noise, the beam
can be perfectly deconvolved.}); 2)
the former plus instrumental noise at $5\;\mu$K/beam; and 3) the
previous plus point sources below a flux limit of $100\;\mu$Jy at
15~GHz.  In this last case, we are assuming that all sources brighter
than the flux limit are explicitly subtracted; for example, both AMI
and the SZA\footnote{\tt http://astro.uchicago.edu/sze} plan long
baseline observations for point source removal.

\begin{figure}
\includegraphics[scale=0.5]{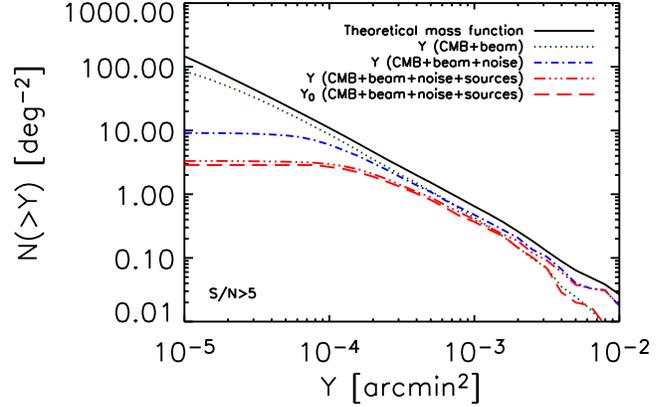}\\
\caption{Cluster counts in terms of integrated $Y$ 
for the input concordance model (black solid line) and for detected
clusters: the green dotted line gives the counts neglecting the
effects of instrumental noise and point sources (CMB+beam$=$2~arcmin
FWHM); the blue dash--dotted line includes instrumental noise
(5~$\mu$K/beam); the red dash--triple--dotted line further
includes the effects of residual point sources after explicit
subtraction of all sources with flux greater than $100\; \mu$Jy (see
text).  These are all plotted as functions of the {\em true} total
flux $Y$.  The red dashed line shows the observed counts for the
latter case in terms of the {\em observed} flux $\Yo$.  }
\label{fig:intcounts}
\end{figure}

Integrated source counts in terms of total cluster flux $Y$ (measured
in {\em arcmin}$^2$) are shown in Figure~\ref{fig:intcounts}.  The
theoretical counts for the fiducial model are given by the solid black
line, while the other curves give the counts from our simulated
observations.  They are plotted in terms of {\em true} flux $Y$,
except for the red dashed curve that gives the counts as a function of
{\em observed} flux $\Yo$, as would actually be observed in a survey.
Differences between the detected cluster counts and the theoretical
prediction (black solid line) reflect catalog incompleteness; the
nature of this incompleteness is the focus of our discussion.  The
influence of photometric errors is illustrated by the difference
between the observed counts as a function of observed flux 
(red dashed curve) and the detected--cluster counts given
as a function of true flux.

\begin{figure}
\includegraphics[scale=0.5]{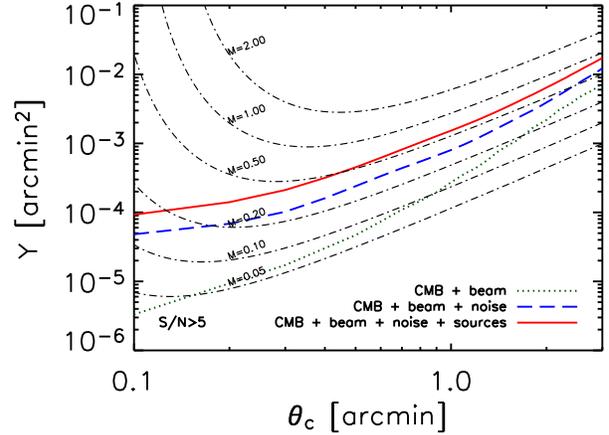}\\
\caption{Selection in the parameter plane of total flux $Y$ and
core radius $\thetac$.  The three curves correspond to
the different simulated cases, as indicated in the legend; all 
correspond to a cut at signal--to--noise of 5.  The dot--dashed
lines in the background give contours of constant mass in this plane;
each is parameterized by redshift $z$.  Note that cluster selection
does not follow a simple flux cut, which would be a horizontal line, 
nor a simple mass cut.  Photometric errors are neglected in this plot, 
meaning that observed cluster parameters $\Yo$ and $\thetac$ equal 
the true values $Y$ and $\thetac$.
}
\label{fig:Ythetacut}
\end{figure}

\subsection{Catalog completeness}

It is important to understand the exact nature of the incompleteness
evident in Figure~\ref{fig:intcounts}, and we shall now demonstrate
that it is not simply a function of total flux.  Our detection
algorithm operates as a cut at fixed signal--to--noise, which
leads to the following constraint on (true) cluster parameters $Y$ and
$\thetac$:
\begin{equation}
Y = y_{\rm est} \int d\Omega\; \tauc(\hat{n})  
\geq \left(\frac{S}{N}\right) \sigma_{\theta_c}\int d\Omega\; \tau_c(\hat{n})
\end{equation}
where $y_{\rm est}$ is the central Compton parameter 
estimated by the filter matched to a cluster of core radius $\thetac$, 
and the filter noise on this scale is given by
Equation~(\ref{eq:filternoise}).  Figure~\ref{fig:Ythetacut} shows the
resulting selection curves for our three cases in the $Y$--$\thetac$
plane at $S/N\geq 5$.  
Note that we are speaking in terms of true cluster parameters,
leaving the effects of photometric errors aside for the moment.

It is clear from this Figure that cluster selection does not
correspond to a simple flux cut -- it depends rather on a combination
of both source flux and angular extent.  The exact form of this
dependence is dictated by the noise power spectrum, which must be
understood to include primary CMB anisotropy.  That this latter
dominates on the larger scales can be seen from the fact that the
three curves approach each other at large core radii.  For smaller
objects, on the other hand, instrumental noise and residual point
source contamination ``pull'' the curve towards higher fluxes relative
to the ideal case that includes only CMB anisotropies (dotted line).

For the solid red curve, we calculate the flux variance induced by
residual point sources at the given filter scale and then add the
equivalent Gaussian noise term to the instrumental noise and CMB
contributions.  One may well ask why the source fluctuations should be
Gaussian given the shallow slope of the radio source counts that would
normally lead to very non--Gaussian statistics.  The fluctuations are
in fact Gaussian, as we have verified with the simulations,
essentially because the source subtraction is performed at higher
angular resolution than the smallest filter scale; in effect, we have
cleaned ``below'' the filter confusion limit, so that the number
of sources/filter beam is large and we approach the Gaussian
limit.  This realistically
reflects what will actually be done with interferometers using long
baseline observations for source subtraction.

The dot--dashed lines in the background of the Figure represent
contours of constant cluster mass $M(Y,\thetac)$.  They result from
inversion of the $Y(M,z)$ and $\thetac(M,z)$ relations, where we
associate cluster core radius with filter scale.  Note that redshift
varies along each contour, and that we have assumed zero scatter in
the relations so that the inversion is one--to--one.  In reality, of
course, they contain intrinsic scatter, due to cluster physics, as
well as observational scatter induced by photometric errors.  The
position of these mass contours depends on both cluster physics and
the underlying cosmology; we may, for example, displace the contours
by changing the gas--mass fraction.  The selection curves, in contrast,
are independent of cosmology and cluster physics, being 
based on purely observational quantities.

Observed clusters populate this plane according to the distribution
$d\No/d\Yo d\thetaco$, which depends on cluster physics, cosmology
and photometry; eq.~(\ref{eq:obs2Mz}) gives it in terms
of the key theoretical quantity, the mass function.  If photometric
errors are assumed to be unimportant, then Eq.~(\ref{eq:nodispersion})
applies and we see that the function $\chi(Y,\thetac)$ is a step
function taking the value of unity above the selection curves, 
and zero below; photometric errors simply ``smooth'' the 
selection function $\Phi$ as manifest by Eq.~(\ref{eq:sfsep}).
Completeness expressed in terms of the function $\chi$ is 
therefore {\em independent} of cluster physics and cosmology.
A more common way to express completeness is by the ratio of detected
to actual clusters as a function of total flux (or angular scale).  At
a given flux, for example, this ratio is the fraction of clusters
falling above the selection curve.  Clearly, it depends on the
distribution of clusters over the plane and is, hence, {\em dependent}
on cluster physics, cosmology and photometry.  We conclude that the
function $\chi$ is a more useful description of a survey.

\begin{figure}
\includegraphics[scale=0.5]{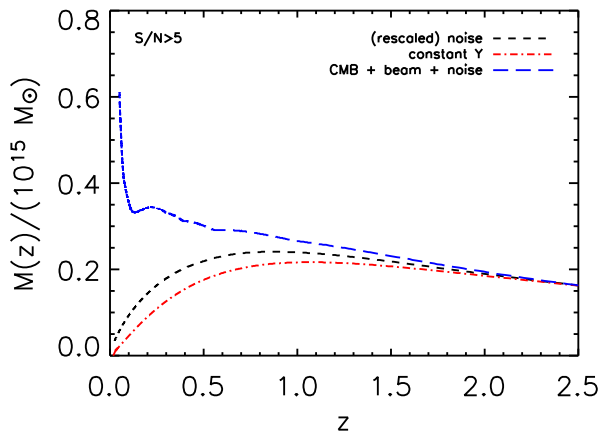}\\
\caption{Detection mass as a function of redshift.  The 
blue long--dashed line shows the result for the case CMB+noise (blue
long--dashed line in Figure~\ref{fig:Ythetacut}).  The rise at low
redshift is due confusion with primary CMB fluctuations that is more
important for nearby clusters with large angular extent. The red
dot--dashed line gives the result for a pure flux--limited catalog
(see text), and the black short dashed line that for observations
without CMB confusion (e.g., multi-frequency).  Relative to a
pure flux--limited catalog, both observed catalogs loose clusters
over a range of redshifts.
}
\label{fig:Mdet}
\end{figure}

Figure~\ref{fig:Ythetacut} provides a concise and instructive view of
cluster selection over the observational plane.  We are of course
ultimately interested in the kinds of objects that can be detected as a
function of redshift, and to this end it is useful to study the {\em
detection mass} shown in Figure~\ref{fig:Mdet}.  This is defined as
the smallest mass cluster detectable at each redshift given the
detection criteria.  For the figure, we assume that there is no
scatter in the $\Yo(M,z)$ and $\thetaco(M,z)$ relations so that a
selection curve in the observational plane uniquely defines the
function $\Mdet(z)$.  Note that, as emphasized above, these detection
mass curves depend on the assumed cosmology.

We compare three situations in the figure. The blue long--dashed line
gives the detection mass for the case CMB+noise (single frequency
experiment), while the red dot--dashed line shows the result for a
pure flux--limited catalog.  The chosen flux cut corresponds to the
left--most point on the blue long--dashed selection curve in
Figure~\ref{fig:Ythetacut} (CMB+noise).  Finally, the black
short--dashed line gives the detection mass for a case with just
instrumental noise (with the same beam as the previous cases) and no
primary CMB; this approximates the situation for a multi--frequency
experiment which eliminates CMB confusion.  The noise level has been
adjusted such that the selection curve in the $(Y,\thetac)$--plane
matches the previous two cases on the smallest scales. With this
choice, all three detection mass curves overlap at high $z$ as seen
in Figure~\ref{fig:Mdet}. 

\begin{figure}
\includegraphics[scale=0.5]{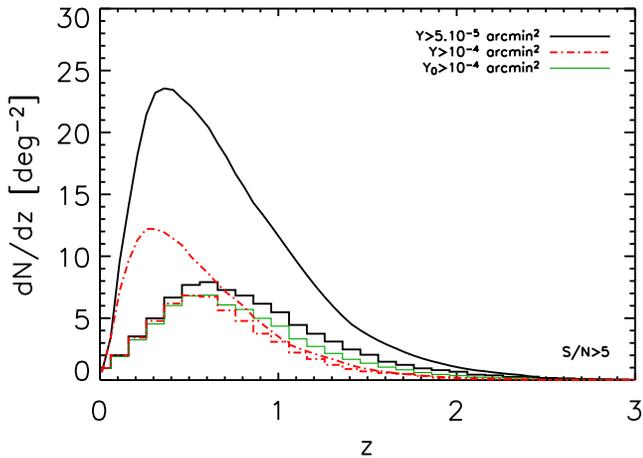}\\
\caption{Redshift distribution of SZ clusters (case 2 -- without
residual point source noise).  The black solid and red dashed curves
give the theoretically predicted counts at the two indicated flux
limits.  Corresponding distributions for the simulated recovered
counts, with the same two flux cuts on the true $Y$, are shown by the
black and red, dashed histograms; the small difference between the two
reflects the flat observed counts in Figure~\ref{fig:intcounts}.  The
lighter, green histogram shows the simulated counts cut at an
{\em observed} flux of $\Yo>10^{-4}$~arcmin$^2$.}
\label{fig:zdists}
\end{figure}

We see that that the observed catalog (blue long--dashed curve) 
looses clusters (i.e., has a higher detection mass) over a 
broad range of redshifts relative to the pure flux--limited catalog
(red dot--dashed line); the effect is most severe for nearby objects, 
whose large angular size submerges them in the primary CMB anisotropies, 
but it remains significant out to redshifts of order unity.  This 
is also reflected in the redshift distribution of Figure~\ref{fig:zdists}
to be discussed below.  We note in addition that even multi--frequency 
experiments loose clusters over a rather broad range of redshifts, as
indicated by the difference between the lower two curves.

Simulations are needed to evaluate the importance of factors not
easily incorporated into the simple analytic calculation of the
cluster selection curve; these include source blending and morphology,
other filtering during data analysis, etc...  Using our simulations,
we find that cluster detection in mock observations closely follows
the analytic predictions, thus indicating that blending does not
significantly change the above conclusions, at least for the case
under study -- a 2~arcmin beam with noise at a level of $5\;\mu$K/beam
-- representative of planned interferometer arrays. As our current
simulations only employ spherical beta model profiles, they only test
for the importance of blending effects; future work will include more
realistic profiles taken, for example, from hydrodynamical N--body
simulations.  The simulations are also crucial for correctly
evaluating the photometric precision of the survey catalog.  Contrary
to the situation for cluster detection, we find that blending greatly
affects photometric measurements: photometric scatter from the
simulations is significantly larger than expected based on the S/N
ratio, whether the threshold is taken at S/N=5 or 3.

\begin{figure}
\includegraphics[scale=0.5]{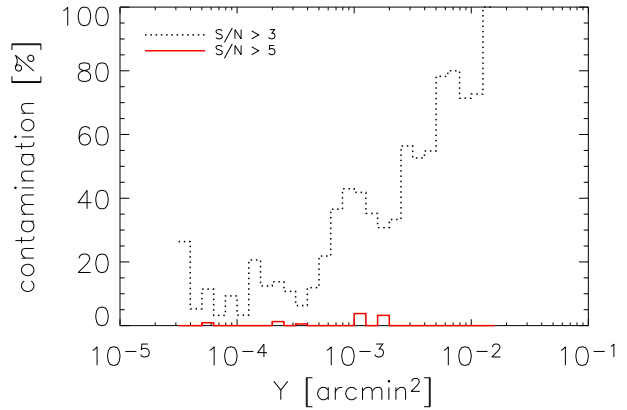}\\
\caption{Contamination rate for a single frequency survey 
as a function of total flux for two different detection thresholds.
The histograms give the percentage of sources that are false
detections in catalogs extracted from our simulations.}
\label{fig:contam}
\end{figure}

\subsection{Catalog contamination}

Contamination by false detections is a separate function that can only
be given in terms of observed flux and angular (or filter) scale; once
again, simulations are crucial for evaluating effects such as blending
and confusion.  Figure~\ref{fig:contam} shows the contamination level
in our extracted catalogs as a function of total flux $Y$.  The level
is significantly higher than expected from the S/N ratio, indicating
that confusion and blending effects are clearly important.  This is
most obvious for the case with S/N=3, where contamination rises
towards the high flux end due to confusion with primary CMB
fluctuations that are more prevalent on larger angular scales.  Even
at relatively low flux levels around $10^{-4}$~arcmin$^2$, we see that the
contamination rate remains near or above 10\% for the S/N=3 case.
This quantifies the the expectation that single frequency surveys
will contend with a non--negligible level of contamination.

\subsection{The redshift distribution}

The example of extracting cosmological constraints from the redshift
distribution of SZ detected clusters affords a good illustration of
the importance of understanding the selection function.  These
constraints arise from the shape of the cluster redshift distribution,
which is affected by such parameters as the matter density
(\cite{oukblan97}) and the dark energy equation--of--state
(\cite{wang98}); this is in fact one of the primary motivations for
performing SZ cluster surveys (\cite{haiman01}).  The important point is
that the redshift distribution expected in a given cosmological model
also depends on the catalog selection function.  In the following
discussion, we assume that the $Y(M,z)$ and $\thetac(M,z)$ relations
are perfectly known.

Consider the redshift distributions shown in Figure~\ref{fig:zdists}
for an observation where residual point source contamination has been
reduced to a negligible level (case 2).  The black line represents the
theoretical distribution for clusters with total flux $Y>5\times
10^{-5}$~arcmin$^2$, which corresponds to the point source detection
limit on the smallest filter scale (leftmost point on the dashed blue
curve in Figure~\ref{fig:Ythetacut}).  This predicted distribution is
very different from the actual distribution of clusters shown as the
black histogram.  It is clearly impossible to deduce the correct
cosmological parameters by fitting a flux--limited theoretical curve
to the observed distribution.  This demonstrates that the
point--source flux limit cannot be used to model the catalog redshift
distribution, which is already clear from the fact that the counts in
Figure~\ref{fig:intcounts} have already turned over and the catalog is
clearly incomplete.

One can try to cut the catalog at a higher flux limit of
$Y>10^{-4}$~arcmin$^2$, where the observed counts just begin to
flatten out and incompleteness is not yet severe. Comparison of the
dashed red line -- theoretically predicted counts at this flux limit
-- with the red dashed histogram shows that the observed distribution
still differs significantly from the predicted flux--limited redshift
distribution.  Modeling the observed catalog as a pure flux cut would
again lead to incorrect cosmological constraints. In order to extract
unbiased parameter estimates, one must adequately incorporate the
full catalog selection criteria.  

\begin{figure}
\includegraphics[scale=0.5]{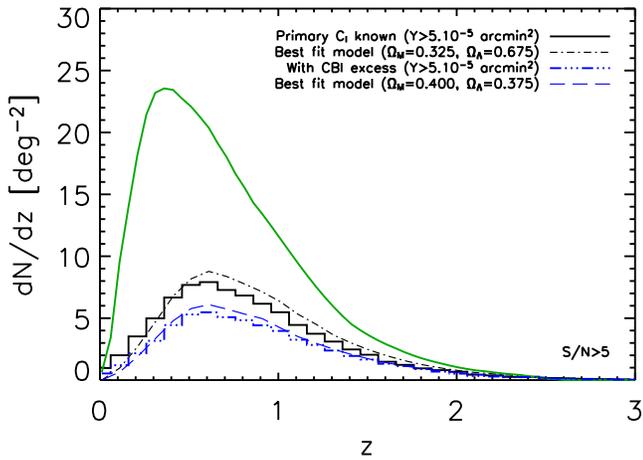}\\
\caption{Effect of incorrect modeling of the selection function.  The 
black continuous curve and black (upper) histogram repeat the results
of Figure~\ref{fig:zdists} for catalogs cut at a flux of $Y=5\times
10^{-5}$~arcmin$^2$ -- the former for a pure flux--limited catalog,
the latter for the clusters extracted from our concordance model
simulations with the expected primary CMB power spectrum
[$(\OmM,\OmL)=(0.3,0.7)$]; note that the histogram is calculated as
the average over 50 simulations of a 3$\times3$ square degree survey
field.  The light black, dot--dashed curve is the best--fit model to
the redshift distribution from a single such simulation; the
constraints from for this fit are shown in
Fig.~\ref{fig:const_cbi}. The lower (blue) histogram shows the
distribution of clusters extracted from the same 50 simulations, but
with excess primary CMB power added at high $l$ (see text); once
again, the histrogram is the average over the ensemble of simulations.
The blue dashed curve shows the best--fit for the same realization as
before -- but now with the excess -- {\em when ignoring the excess in
the fitting} (incorrect selection function modeling).  Corresponding
constraints are shown in Fig.~\ref{fig:const_cbi}.  Both fits
are statistically acceptable (see text).
}
\label{fig:zdist_cbi}
\end{figure}

We may illustrate this point by considering the effect of an
un--modeled CMB power excess at high $l$, such as suggested by the CBI
experiment (\cite{mason01}).  As we have seen in
Figure~\ref{fig:Ythetacut}, the primary CMB fluctuations influence the
exact form of the selection curve in the $(Y,\thetac)$ plane; their
power on cluster scales must therefore be accurately known to
correctly model the cluster selection function.  The black curve and
black histogram in Figure~\ref{fig:zdist_cbi} repeat the results of
Figure~\ref{fig:zdists} for a cut at $Y>5\times 10^{-5}$~arcmin$^2$.
In particular, the black histogram gives the redshift distribution of
clusters extracted from simulations including a CMB power spectrum
corresponding to the concordance model.  The blue (lower) histogram
shows the redshift distribution for clusters extracted from
simulations in which additional CMB power has been added at high $l$
-- a constant power of $l(l+1)C_l/2\pi = 20\; \mu$K was smoothly
joined to the concordance model CMB spectrum (just below $l=2000$) and
continuing out to $l=3000$.  Instead of plunging towards zero, as
expected of the primary CMB fluctuations in the concordance model,
this second model levels off at a constant power level on cluster
scales. This has an important effect on cluster detection, as clearly
evinced in the Figure.  

We now examine the effect of ignoring this excess power in an analysis
aimed at constraining cosmological parameters.  This means that we
ignore the excess both in the construction of the matched filter and
in the selection function model needed for the fit.  The former has
only a relatively minor effect on the catalog extraction and observed
histogram.
The second effect is much more serious, as we now demonstrate.

\begin{figure}
\includegraphics[scale=0.5]{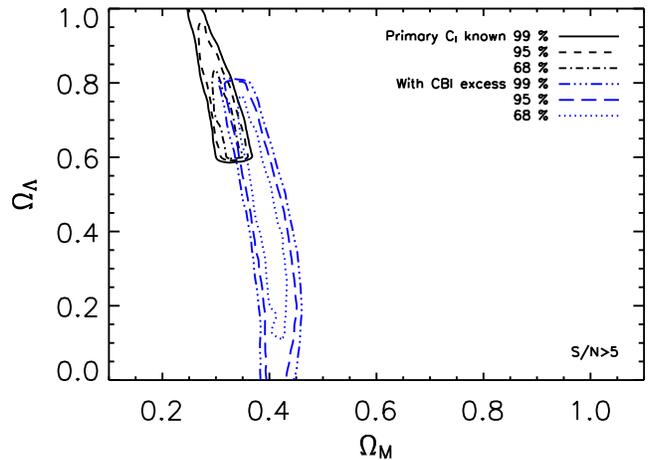}\\
\caption{Confidence contours for the fits discussed in 
Figure~\ref{fig:zdist_cbi}, shown for a survey covering 3$\times$ 3
sq. degrees.  The upper (black) contours correspond to the case where
the selection function is correctly modeled (no excess CMB power at
high $l$); the best--fit parameters are $(\OmM, \OmL) = (0.325,0.675)$
and $1\sigma$ contours fully enclose the true (simulation input)
cosmological values of $(0.3,0.7)$.  The larger (blue) contours
represent the situation when the CMB excess is not properly accounted for
by the selection function model.  The best--fit parameter values are
significantly biased -- $(0.4,0.375)$ -- and the true parameter values,
lie outside the 99\% contour.  In both cases the fits are acceptable
(see text).  }
\label{fig:const_cbi}
\end{figure}

Consider constraints on the parameter pair $(\OmM,\OmL)$ by fitting
models to the redshift distribution of a 3$\times$3 square degree
survey.  Note that the histograms shown in the figures are in fact
averages taken over an ensemble of 50 such simulations, to avoid
confusing statistical fluctuations.  For the present example, however,
we fit models to the redshift distribution from a single simulation.
During the fit, we fix the Hubble parameter to its standard value
($\Ho=70~$km/s/Mpc) and adjust the power spectrum normalization
$\sigma_8$ to maintain the observed present--day cluster abundance
(following \cite{pierp01}).  For our simplified case of zero--scatter
relations between $(\Yo,\thetaco)$ and $(M,z)$, both the selection
function $\Phi$ and the intrinsic scatter function $T$ contain Dirac
delta functions that collapse the various integrals in
Eqs.~\ref{eq:obs2Mz} and \ref{eq:psifunc}.  We then obtain the
following expression for the redshift distribution of observed
clusters brighter than a flux of $\Yo$:
\begin{equation}\label{eq:modelfit}
\frac{d\No}{dz}(>\Yo) = \int_{M(\Yo,z)}^\infty 
	dM\; \chi\left[Y(M,z),\thetac(M,z)\right] \frac{dN}{dzdM}
\end{equation}
where $M(Y,z)$ is the zero--scatter relation between flux and
mass and redshift.  All selection effects are encapsulated in the
completeness function $\chi$, whose dependence on the primary
CMB power is the focus of our present discussion. 

We consider two cases: the first with the expected concordance primary
CMB power spectrum, the second with the CBI--like excess power.  In
the first case, we adopt the true power spectrum for catalog
construction and modeling of $\chi$ -- the selection function is properly
modeled.  In the second situation, we ignore the excess in both
catalog construction and in fitting -- the selection function is
incorreclty modeled.  When correctly modeling the selection function,
we find best--fit values of $(\OmM,\OmL)=(0.325,0.675)$.  The light
black dot--dashed curve in Figure~\ref{fig:zdist_cbi} shows that this
model reasonably reproduces the predicted redshift distribution (black
solid histogram), and the $1\sigma$ contours in
Figure~\ref{fig:const_cbi} enclose the true (simulation input) values.
The fit is good with a reduced $\chi^2=0.94$ (34
degrees--of--freedom).  When incorrectly modeling the selection
function, on the other hand, we find biased best--fit values of $(0.4,0.375)$,
and, as shown in Figure~\ref{fig:const_cbi}, the true parameter values
fall outside the 99\% confidence contours.  Furthermore, this biased
fit is acceptable with a reduced $\chi^2=1.17$ (31
degrees--of--freedom), giving no indication of its incorrectness.  The
redshift distribution of this model is shown as the light dashed
(blue) curve in Fig.~\ref{fig:zdist_cbi}, faithfully reproducing the 
(averaged) histogram for this case.  This is a particularly telling
example of the importance of the selection function, because the
primary CMB power on cluster scales is at present not well known.  It
will have to be constrained by the same experiments performing SZ
cluster surveys; cosmological constraints will be correspondingly
degraded, a subject we return to in a future work.

For another example of incorrect modeling of the selection function,
consider that $\beta$ and $\thetac$ of real clusters may not behave
as we assume when constructing the matched filter.  This will
bias flux measurments and displace the selection curve in 
the $(Y,\thetac)$ plane relative to our expectations, leading
to an incorrect selection function model.  As above, this will
yield biased parameter estimates.

As a final note, and returning to Figure~\ref{fig:zdists}, we show the
distribution of detected clusters at the higher flux cut as a function
of {\em observed} flux with the lighter, green histogram.  The
difference with respect to the corresponding distribution in terms of
true flux (the red, dashed histogram) reflects statistical photometric
errors; note that in fact this tends to falsely increase the number of
objects seen at the higher redshifts.  Although in this case
photometric errors are of secondary importance to the observed
redshift distribution (completeness effects dominate), they must also
be fully accounted for in any cosmological analysis.

\section{Discussion and Conclusions}
\label{sec:dis}
Our aim as been to emphasize the importance of understanding the SZ
cluster selection function, as for any astronomical survey. We
proposed a general definition of the selection function that can be
used to directly relate theoretical cluster distributions to observed
ones, and which has the nice property of clearly separating the
influence of catalog incompleteness and photometric errors.  It is a
function of both observing conditions and of the detection and
photometry algorithms used to construct the survey catalog.  Defined
over the (true) total flux--angular size plane, however, the selection
function is independent of cosmology and cluster physics; its
connection to theoretical cluster descriptors, such as mass and
redshift, on the other hand, depends on both.  A common way of 
quoting incompleteness in terms of total flux is similarly 
sensitive to cluster physics and underlying cosmology.

Using a matched spatial filter (Melin et al. 2004), we studied the
selection function for single frequency SZ surveys, such as will be
performed with upcoming interferometers\footnote{Although we have not
here modeled the actual data taking in the visibility plane.}.  Our 
main result is that a SZ catalog is not simply flux
limited, and this has implications for cosmological studies.  
A simple analytic argument shows the exact manner in which
catalog selection depends on both cluster flux and angular size;
simulated observations indicate that this simple estimate is quite
accurate and little affected by blending, although future work needs
to take into account more realistic cluster profiles.  We also noted
that noise induced by residual point sources tends to be Gaussian,
because subtraction of the brightest sources will be done at higher
angular resolution than the smallest filter scale in the SZ maps.

The implications for cosmological studies were illustrated with the
redshift distribution, which will serve to constrain cosmological
parameters in future surveys.  Theoretical redshift distributions
based on a simple flux limit cannot fit observed distributions; at
best they would lead to biased estimates of cosmological parameters.
One must incorporate the complete selection criteria depending on both
flux and angular extent, and hence have a good understanding of the
catalog selection function.  This understanding depends on a number of
astrophysical factors in addition to instrumental parameters.  Our
example of an unmodeled primary CMB power excess (relative to the
adopted concordance model) on small angular scales ($l\geq 2000$)
highlights the point: we obtained biased parameter estimates because
the selection function was incorrectly modeled; note that the false
fit was in fact a good fit to the data, according to the $\chi^2$.
Other factors, for example, cluster morphology and its potential
evolution, will also play a role.  In the particular case of the CMB
power excess, we note that accurate knowledge of the primary CMB power
on cluster scales will come from the same experiments performing the
cluster surveys.  It will be necessary to constrain the primary CMB
power at the same time as cluster extraction, a point we return to in
a future work.

An issue currently receiving attention in the literature concerns SZ
survey ``calibration'', by which is meant the empirical Establishment
of the $Y(M,z)$ relation.  This is clearly essential for any
cosmological study.  The fact that SZ catalog selection depends not
only on total flux but also on angular size complicates the question
of survey calibration, for it implies that one must additionally
establish a $\thetac(M,z)$ relation, or its equivalent with some other
angular size measure.  In fact, since the dispersion on $Y$ and
$\thetac$ will in general be correlated, we need the full joint
distribution for these observables as a function of mass and redshift.
Photometric errors, which we find can be significant, further
complicate the issue by increasing scatter in observed relations and
hence making them more difficult to obtain.

Although in this work we have focused our detailed study on single
frequency surveys, the general conclusions should carry over to
multiple frequency observations.  In closing we note that the
selection function obviously has equally important implications for
other studies based on SZ--detected cluster catalogs, such as spatial
clustering, etc...  For many of these studies, photometric errors, which
we have only briefly touched on here, will take on even greater
importance.

\begin{acknowledgements}
J.--B. Melin and J. G. Bartlett thank the France Berkeley Fund (grant
``Precision Cosmology from CMB analysis'') and the Lawrence Berkeley
Laboratory for financial assistance and hospitality for a visit during
which part of this work was completed.  We also thank our anonymous referee
for helpful and thoughfull comments. 

\end{acknowledgements}

\noindent Web pages of various SZ experiments:
\begin{itemize}{\small
\item ACBAR {\tt http://cosmology.berkeley.edu/group/swlh/acbar/}
\item ACT {\tt http://www.hep.upenn.edu/$\sim$angelica/act/act.html}
\item AMI {\tt http://www.mrao.cam.ac.uk/telescopes/ami/index.html}
\item AMiBA {\tt http://www.asiaa.sinica.edu.tw/amiba}
\item APEX {\tt http://bolo.berkeley.edu/apexsz}
\item BOLOCAM {\tt http://astro.caltech.edu/$\sim$lgg/bolocam\_front.htm}
\item SPT {\tt http://astro.uchicago.edu/spt/}
\item SZA {\tt http://astro.uchicago.edu/sze}
\item Planck {\tt http://astro.estec.esa.nl/Planck/}}
\end{itemize}

\end{document}